\documentclass[fleqn,usenatbib]{mnras}

\usepackage{newtxtext,newtxmath}
\usepackage[T1]{fontenc}

\DeclareRobustCommand{\VAN}[3]{#2}
\let\VANthebibliography\thebibliography
\def\thebibliography{\DeclareRobustCommand{\VAN}[3]{##3}\VANthebibliography}

\usepackage{subcaption}
\usepackage{graphicx}	% Including figure files
\usepackage{gensymb}    % Degree symbol
\usepackage{amsmath}	% Advanced maths commands
\usepackage{physics}
\usepackage{comment}
\usepackage{fancyvrb} % For inline verbatim text(monospaced font for code names)  (RB)
\newcommand{\tenud}[3]{#1^{#2}_{\hphantom{#2}#3}}

\newcommand{\liedv}[1]{\mathscr{L}_{#1}} 	% Lie Derivative
\newcommand{\flatdv}[1]{\partial_{#1}} 	% Partial derivative
\newcommand{\codv}[1]{\nabla_{#1}} 		% Covariant Derivative
\newcommand{\condv}[1]{\nabla^{#1}} 		% Contravariant Derivative
\newcommand{\gb}{\beta}
\newcommand{\gp}{\psi}
\newcommand{\gt}{\theta}
\newcommand{\gf}{\phi}
\newcommand{\gF}{\Phi}
\newcommand{\gm}{\mu}

\newcommand{\gl}{\lambda}

\newcommand{\gD}{\Delta}

\newcommand{\gr}{\rho}
\newcommand{\paren}[1]{\left(#1\right)}
\newcommand{\colch}[1]{\left[#1\right]}

\title[Magnetically Confined Mountains on Accreting Neutron Stars in GR]{Magnetically Confined Mountains on Accreting Neutron Stars in General Relativity}

\author[P. H. B. Rossetto et al.]{
Pedro H. B. Rossetto,$^{1}$\thanks{E-mail: phbrossetto@gmail.com}
Jörg Frauendiener,$^{1}$
Ryan Brunet$^{2,3}$
and Andrew Melatos$^{2,3}$
\\
$^{1}$Department of Mathematics \& Statistics, University of Otago, 730 Cumberland Street, Dunedin 9016, New Zealand
\\
$^{2}$School of Physics, University of Melbourne, Parkville, VIC 3010, Australia
\\
$^{3}$ARC Centre of Excellence for Gravitational Wave Discovery (OzGrav), University of Melbourne, Parkville, VIC 3010, Australia
}

\date{Accepted XXX. Received YYY; in original form ZZZ}

\pubyear{2022}

\begin{document}
\label{firstpage}
\pagerange{\pageref{firstpage}--\pageref{lastpage}}
\maketitle

\begin{abstract}
The general relativistic formulation of the problem of magnetically confined mountains on neutron stars is presented, and the resulting equations are solved numerically, generalising previous Newtonian calculations. The hydromagnetic structure of the accreted matter and the subsequent magnetic burial of the star's magnetic dipole moment are computed. Overall, it is observed that relativistic corrections reduce the hydromagnetic deformation associated with the mountain. The magnetic field lines are curved more gently than in previous calculations, and the screening of the dipole moment is reduced. Quantitatively, it is found that the dimensionless dipole moment ($m_{\rm d}$) depends on the accreted mass ($M_{\rm a}$) as $m_{\rm d} = -3.2\times10^{3}M_{\rm a}/M_\odot + 1.0$, implying approximately three times less screening compared to the Newtonian theory. Additionally, the characteristic scale height of the mountain, governing the gradients of quantities like pressure, density, and magnetic field strength, reduces by approximately $40\%$ for an isothermal equation of state.
\end{abstract}

\begin{keywords}
accretion -- magnetic fields -- MHD -- relativistic processes -- stars: neutron
\end{keywords}

%%%%%%%%%%%%%%%%%%%%%%%%%%%%%%%%%%%%%%%%%%%%%%%%%%

%%%%%%%%%%%%%%%%% BODY OF PAPER %%%%%%%%%%%%%%%%%%

\section{Introduction}
\label{sec:introduction}
Neutron stars with nonaxisymmetric mountains are sources of continuous gravitational waves (CWs) \citep{zimmermann1979Gravitational, sieniawska2019Continuous,riles2023Searches}. Therefore, searches for CW signals from neutron stars have targeted a diverse range of astrophysical scenarios, including likely neutron stars in supernova remnants \citep{2021ApJ...921...80A, 2022PhRvD.105h2005A}, known radio and X-ray pulsars \citep{2020ApJ...902L..21A,2021ApJ...913L..27A,2021ApJ...922...71A, 2022PhRvD.105b2002A, 2022ApJ...932..133A, 2022ApJ...935....1A}, and undiscovered neutron stars via all-sky surveys \citep{2021PhRvD.103f4017A, 2021PhRvD.104l2004A, 2022ApJ...929L..19C, 2022PhRvD.105j2001A}. 

One plausible mechanism for forming mountains on accreting neutron stars is polar magnetic burial \citep{uchida1981Equilibrium,hameury1983Magnetohydrostatics,cheng1998Magnetic,pm04}. In this process, mass from the inner edge of the accretion disk follows the magnetic field lines and lands on the magnetic poles of the star. This results in a reduction of the star's magnetic dipole moment and in an increase of its mass quadrupole moment. The latter outcomes are consistent respectively with the magnetic field distribution observed in recycled pulsars \citep{wijers1997Evidence} and the spin period distribution observed in low-mass X-ray binaries (LMXBs) \citep{chakrabarty2003Nuclearpowered}.

\cite{pm04} have considered the self-consistent equilibrium of the polar magnetic mountain by using the fact that matter does not move through magnetic flux surfaces. They showed that mountains bury the magnetic field and reduce the magnetic dipole moment of the star. This may be one explanation for the low dipole moments observed for recycled pulsars \citep{wijers1997Evidence}. \cite{melatos2005Gravitational} considered the implications of these results for the quadrupole moment of the star and, therefore, the emission of gravitational radiation. This can be linked with the slow rotation of neutron stars found in accreting systems \citep{chakrabarty2003Nuclearpowered}.

Several aspects of mountain formation by polar magnetic burial have been studied in the literature, including hydromagnetic stability \citep{vigelius2008Threedimensional, mukherjee2012Phasedependent, mukherjee2013MHD}, ohmic and thermal relaxation \citep{vigelius2009Resistive, suvorov2019Relaxation}, superconductivity \citep{passamonti2014Quasiperiodic,sur2021Impact}, crustal sinking \citep{choudhuri2002Diamagnetic,wette2010Sinking}, equations of state \citep{priymak2011Quadrupole,mukherjee2017Revisiting}, triaxial configurations \citep{singh2020Asymmetric}, toroidal fields and higher magnetic multipole moments \citep{suvorov2020Recycled, fujisawa2022Magnetically}. However, all of the cited studies were done assuming Newtonian gravity. The aim of the present paper is to formulate the problem of polar magnetic burial and mountain formation in general relativity and compare it with previous Newtonian calculations. The paper has the following structure. In Section \ref{sec:GRMHD}, we construct the relativistic theory of self-consistent magnetically confined mountains on neutron stars including mass-flux conservation. In Section \ref{sec:numerics}, we describe the numerical method used to solve the equations in Section \ref{sec:GRMHD}. In Section \ref{sec:results}, we investigate the hydromagnetic structure of the system by calculating the mountain's equilibrium mass density, the distortion of the magnetic field lines, and the burial of the star's magnetic dipole moment. We also compare our results with the Newtonian ones. Finally, in Section \ref{sec:conclusion}, we review the main findings of the paper, the limitations of our analysis and possible extensions of the current work.

Even though magnetically accreted mountains have only been considered in Newtonian gravity, magnetically deformed mountains of isolated pulsars have already been considered in relativity \citep{bonazzola1996Gravitational, chatterjee2021Structure, colaiuda2008Relativistic, konno1999Deformation, mallick2014Deformation, sengupta1998Evolution, zamani2021Analytic}. Most of the aforementioned papers model the deformation using general relativistic magnetohydrodynamics (GRMHD) axisymmetric equilibrium, for which a full general covariant approach can be found in \cite{gourgoulhon2011magnetohydrodynamics}.

\section{Hydromagnetic mountain equilibrium in general relativity}
\label{sec:GRMHD}
The equilibrium structure of a mountain formed by polar magnetic burial is governed by two physical laws: hydromagnetic force balance, expressed through the Grad-Shafranov equation and discussed in Section \ref{sec:theory}, and magnetic flux-freezing, expressed through an integral constraint on the mass-flux ratio and discussed in Section \ref{sec:flux-freezing}. These laws are supplemented by boundary conditions, which are specified and justified in Section \ref{sec:bc}.

\subsection{Grad-Shafranov equation}
\label{sec:theory}

We work in traditional Schwarzschild coordinates $(t,r,\theta,\phi)$. The accreted mass $M_{\rm a}$ satisfies $10^{-6} \lesssim M_{\rm a} / M_\odot \lesssim 3\times10^{-5}$ for the purposes of the calculations in this paper. This mass range is consistent with the early stages of accretion in LMXBs \citep{taam1986Magnetic}. We have chosen the upper limit to be deliberately lower than the realistic astrophysical maximum to avoid magnetic bubbles \citep{pm04, payne2007Burial}. The formation and evolution of magnetic bubbles in the relativistic context will be studied in a future paper. Accordingly, as a first pass at the problem, it is reasonable to neglect the contribution of $M_{\rm a}$ to the spacetime geometry and assume that accretion occurs in a Schwarzschild background generated by a neutron star with gravitational mass $M_\ast \gg M_{\rm a}$. Then the metric is given by\footnote{As is common in the relativistic literature, we work in a geometrized unit system where $G=c=1$. At times, when comparisons with the Newtonian formalism are necessary, we restore the $G$ and $c$ by dimensional analysis. We use the Gaussian unit system for electromagnetic quantities, and the magnetic field has a dividing factor of $\sqrt{4\pi}$.}
\begin{equation}
    \label{metric}
    \dd{s}^2=-e^{2\Phi}\dd{t}^2+e^{-2\Phi}\dd{r}^2+r^2\dd{\theta}^2+r^2\sin^2\theta\dd{\phi}^2,
\end{equation}
with
\begin{equation}
	\label{schwarz_phi}
	\gF(r)=\frac{1}{2}\ln(1-\frac{2M_*}{r}).
\end{equation}

In component notation, the equations of motion take the form
\begin{align}
    \label{cont_eqn}
    &\codv{a}(\gr u^a)=0
    \\
    &\label{mom_eqn}
    \codv{a}T^{ab}=0,
\end{align}
where $\nabla_a$ denotes the covariant derivative and $\gr$, $u^a$ and $T^{ab}$ are respectively the rest-mass density, the four-velocity and the energy-momentum tensor. The latter quantity can be split into its hydrodynamic and electromagnetic parts, viz. $T^{ab} = T^{ab}_{\rm f} + T^{ab}_{\rm em}$, with
\begin{equation}
    \label{T_fluid}
    T^{ab}_{\rm f}=\gr hu^a u^b+g^{ab}p
\end{equation}
and
\begin{equation}
    \label{T_em}
    T^{ab}_{\rm em}=F^{ad}\tenud{F}{b}{d}-\frac{1}{4}g^{ab}F^{de}F_{de}.
\end{equation}
In equations \eqref{T_fluid} and \eqref{T_em}, $p$ is the pressure, $F_{ab}$ is Faraday's electromagnetic tensor,
\begin{equation}
    \label{enthalpy}
    h=\frac{e+p}{\gr}
\end{equation}
is the specific enthalpy, and $e$ is the energy density. The four-velocity $u^a$ is a time-like unit vector with $u_au^a=-1$ according to our conventions.

The Faraday tensor can be decomposed into the electric and magnetic fields, namely the covectors $E$ and $B$, measured by an observer with four-velocity $u$ as
\begin{equation}
    \label{faraday_decomp}
    F=u\wedge E+\star(u\wedge B),
\end{equation}
where \eqref{faraday_decomp} is written in terms of differential forms, and $\star$ and $\wedge$ denote the Hodge dual and wedge product respectively. In equation \eqref{faraday_decomp} we also use that $E$ and $B$ are defined to be perpendicular to $u$. Conversely, given $F$, we can reconstruct the electric and magnetic fields via
\begin{align}
    \label{E_from_F}
    E_a&=F_{ab}u^b,
    \\
    \label{B_from_F}
    B_a&=(\star F)_{ab}u^b.
\end{align}
$F$ satisfies Maxwell's equations
\begin{align}
    \label{max_eqn1}
    \dd{F}&=0
    \\
    \label{max_eqn2}
    \dd{(\star F)}&=\star J,
\end{align}
where $J$ is the current density 4-vector.

We model an accreted mountain, which is static and cylindrically symmetric about the magnetic axis \citep{pm04}. In reality, accretion violates axisymmetry on the free-fall time-scale, e.g.\ via transient finger-like flows formed through the Rayleigh-Taylor instability at the disk-magnetosphere boundary \citep{romanova2008Unstable,romanova2015Accretion}. However, we stick with axisymmetry in this paper as a first pass to keep the focus on the new effects introduced by general relativity and to facilitate comparison with previous axisymmetric calculations \citep{melatos2001Hydromagnetic,pm04}. A fuller treatment is achieved best through time-dependent, three-dimensional, magnetohydrodynamic simulations \citep{romanova2015Accretion}, which track nonaxisymmetric processes over many stellar rotations, cf. \citet{basko1975Radiative}. As the metric \eqref{metric} is also static and axisymmetric, any tensorial quantity related to spacetime, fluid, or electromagnetic fields must have vanishing Lie derivatives with respect to the Killing vectors $e^{a}_{t}=\flatdv{t}^a$ and $e^{a}_{\gf}=\flatdv{\gf}^a$. If the fluid is a perfect conductor, then the electric field in the fluid's frame of reference vanishes, with $E_a=F_{ab}u^b=0$ and $u^a=e^{-\gF}e_{t}^{a}$. Assuming furthermore that the magnetic field is purely poloidal, i.e. $B_ae_{\gf}^{a}=0$, we can write (see Appendix \ref{app:sym})
\begin{equation}
    \label{F_ab}
    F=\dd{\gf}\wedge\dd{\gp}.
\end{equation}
The function $\gp$ in \eqref{F_ab} is called the magnetic flux function. Thus, we reduce the degrees of freedom in $F_{ab}$ to a single scalar function $\gp=\gp(r,\gt)$.

Upon combining \eqref{mom_eqn}, \eqref{T_em}, and \eqref{F_ab}, and writing out the $r$ and $\theta$ components, we obtain
\begin{equation}
    \label{pre_grad_shaf}
    \gr h\codv{a}\gF+\codv{a}p=\Delta^*\psi\codv{a}\gp,
\end{equation}
where $\Delta^*$ is the Grad-Shafranov operator given by
\begin{equation}
    \label{grad_shaf_op}
    \Delta^*\psi=\frac{1}{\lambda}\paren{\condv{a}\codv{a}\gp-\frac{1}{\gl}\condv{a}\gl\codv{a}\gp},
\end{equation}
and $\lambda=r^2\sin^2\gt$ is the squared length of the axisymmetric Killing vector. Note that \eqref{cont_eqn} is satisfied identically in a static situation. In the Schwarzschild metric \eqref{metric} with \eqref{schwarz_phi}, the operator \eqref{grad_shaf_op} takes the form
\begin{equation}
    \label{grad_shaf_op_coords}
    \gD^*\gp=
    \frac{1}{r^2\sin^2\gt}\colch{\pdv{r}\paren{\paren{1-\frac{2M_*}{r}}\pdv{\gp}{r}}
    +\frac{\sin\gt}{r^2}\pdv{\gt}\paren{\frac{1}{\sin\gt}\pdv{\gp}{\gt}}}.
\end{equation}

Given an equation of state $p=p(e)$, one can find a function $U$ such that $\codv{a}U=(\gr h)^{-1}\codv{a}p$. Then, equation \eqref{pre_grad_shaf} implies that $\gF+U$ is constant along the level surfaces of $\gp$. We write $\gF+U=f(\gp)$ and insert this back into equation \eqref{pre_grad_shaf} to obtain
\begin{equation}
    \label{grad_shaf_eqn}
    \Delta^*\psi=-\rho hf'(\psi),
\end{equation}
where the prime denotes the total derivative with respect to $\psi$. Equation \eqref{grad_shaf_eqn} is the general relativistic Grad-Shafranov equation for magnetohydrodynamic equilibrium. The function $f(\gp)$ is a freely-determined function, when it stands alone in (17). To specify its form uniquely, one must specify the mass between adjacent level surfaces $\psi$ and $\psi+\dd\psi$ for all $\psi$. This translates into an integral constraint on $\psi$, which is described in Section \ref{sec:flux-freezing}.

In this paper, we adopt the isothermal equation of state, to facilitate comparison with \cite{pm04} and to maintain the focus on the new effects introduced by general relativity. The results can be generalized to other equations of state using known methods \citep{priymak2011Quadrupole}. In the relativistic literature \citep{yabushita1973pressure, chavanis2008relativistic}, the isothermal equation of state is given by
\begin{equation}
\label{isot_eos}
p(e)=c_{\rm s}^2e,
\end{equation}
where $c_{\rm s}$ is the sound speed in the accreted plasma. In equation \eqref{isot_eos}, $e$ substitutes for the Newtonian $\rho$ \citep{pm04}. The substitution makes sense from a physical point of view because it preserves physical features arising from a Newtonian isothermal equation of state, such as the constant phase speed of plasma wave modes \citep{rezzolla2013relativistic}.

Equation \eqref{isot_eos} implies
\begin{equation}
    \label{iso_p}
    p=F(\gp)\exp[-(1+c_{\rm s}^{-2})(\gF-\gF_0)],
\end{equation}
where $\Phi_0$ is a reference potential, and hence
\begin{equation}
    \label{iso_grad_shaf_eqn}
    \gD^*\gp=-F'(\gp)\exp[-(1+c_{\rm s}^{-2})(\gF-\gF_0)].
\end{equation}
Equation \eqref{iso_grad_shaf_eqn} is the isothermal general relativistic Grad-Shafranov equation. In the Newtonian limit, equations \eqref{grad_shaf_eqn} and \eqref{iso_grad_shaf_eqn} reduce correctly to equations (6) and (12) in \cite{pm04}. In \eqref{grad_shaf_eqn} and \eqref{iso_grad_shaf_eqn}, $F(\psi)$ is a free function related to $f(\psi)$ which replaces $f(\psi)$ and is specified uniquely by the mass-flux constraint in Section \ref{sec:flux-freezing}. The height of an isothermal mountain is $\lesssim 1\%$ of the stellar radius $R_\ast$ \citep{pm04}, so one can Taylor expand $\Phi(r)$ at the surface to obtain
\begin{equation}
    \label{surf_approx_phi}
    \gF(r)\approx \gF_0+\frac{M_*}{R_*(R_*-2M_*)}(r-R_*),
\end{equation}
with
\begin{equation}
    \label{phi_0}
    \gF_0=\gF(R_*)=\frac{1}{2}\ln(1-\frac{2M_*}{R_*}).
\end{equation}

Using approximation \eqref{surf_approx_phi} we can calculate the pressure (and, therefore, density) scale height. For the typical values $R_*=10^{6}\,\mathrm{cm}$ and $M_*=1.4 M_\odot$, the calculated scale heights are about $40\%$ smaller than the corresponding Newtonian values \citep{pm04}. It is then expected that the relativistic corrections are appreciable, and that they make a mountain smaller.

\subsection{Mass-flux distribution and flux-freezing}
\label{sec:flux-freezing}

If the accreted plasma is perfectly conducting, the system evolves according to the flux-freezing condition of ideal magnetohydrodynamics: charged matter does not cross magnetic flux surfaces, as accretion proceeds. In other words, for all $\psi$, the rest-mass $dM$ between the adjacent, infinitesimally separated level surfaces $\psi$ and $\psi + d\psi$ in equilibrium (i.e.\ after accretion ceases, and the mountain settles) equals the mass added by the accretion process between $\psi$ and $\psi+d\psi$. Flux freezing uniquely determines $F(\psi)$ in \eqref{iso_grad_shaf_eqn}, which must be computed numerically in general. It links $F(\psi)$ self-consistently to the initial conditions and mass-loading history of the accretion process, even though the Grad-Shafranov formalism and equation \eqref{iso_grad_shaf_eqn} itself do not depend on time \citep{pm04,vigelius2008Threedimensional,priymak2011Quadrupole}. 

Equations \eqref{enthalpy}, \eqref{isot_eos} and \eqref{iso_p} must be supplemented by the first law of thermodynamics \citep{rezzolla2013relativistic} 
\begin{equation}
    \label{first_law}
    \dd{e}=h\dd{\rho}
\end{equation}
in order to obtain expressions for the four thermodynamic variables $p$, $e$, $\rho$ and $h$. Then, using equations \eqref{enthalpy}, \eqref{isot_eos}, \eqref{iso_p}, \eqref{first_law} and imposing the correct classical limits (to determine the integration constant), the rest-mass density $\rho$ is given by
\begin{equation}
    \label{rho}
    \gr=\colch{c_{\rm s}^{-2}F(\gp)}^{1/(1+c_{\rm s}^2)}\exp[-c_{\rm s}^{-2}(\gF-\gF_0)].
\end{equation}
We integrate $\rho$ between the flux surfaces $\psi$ and $\psi+d\psi$ by changing to coordinates whose basis vectors are unit vectors perpendicular and tangential to the level surfaces of $\gp$, viz
\begin{equation}
    n^a=\frac{\condv{a}\gp}{|\nabla\gp|},
\end{equation}
and
\begin{equation}
    s^a=\frac{\epsilon^{abc}\codv{b}\gp\codv{c}\gf}{\sqrt{\gl}|\nabla\gp|}
\end{equation}
respectively, where we write $|\nabla\gp|^2=\condv{a}\gp\codv{a}\gp$, and $\epsilon_{abc}=u^d\epsilon_{dabc}$ is the three-dimensional volume form. The exterior derivative of the one-form $s_a=g_{ab}s^b$ vanishes. By Poincare's Lemma, there exists a function $s$ satisfying $s_a=(\dd{s})_{a}$ and hence an infinitesimal volume element
\begin{equation}
    \label{vol_form}
    \dd{V}=\frac{r\sin\gt}{|\nabla\gp|}\dd{s}\dd{\gp}\dd{\gf}.
\end{equation}
Therefore, the rest mass is given by $\int_V\rho\dd{V}$ and the mass-flux ratio is
\begin{equation}
    \label{mass-flux}
    \dv{M}{\gp}=2\pi\int_C \frac{\gr r\sin\gt}{|\codv{} \gp|}\dd{s},
\end{equation}
where $V$ and $C$ denote the volume $r\geq R_*$ and a contour $\psi = {\rm constant}$, respectively.

Substituting equation \eqref{rho} in equation \eqref{mass-flux} and solving for $F(\gp)$ we obtain
\begin{equation}
    \label{F_gp_gr}
    F(\gp)=\paren{\dv{M}{\gp}}^{1+c_{\rm s}^2}\paren{\frac{2\pi}{c_{\rm s}^2}\int_C r\sin\gt|\codv{} \gp|^{-1}e^{-\paren{\gF-\gF_0}/c_{\rm s}^2}\dd{s}}^{-(1+c_{\rm s}^2)}
\end{equation}
using the surface approximation \eqref{surf_approx_phi}. Equation \eqref{F_gp_gr} reduces correctly to equation (14) in \cite{pm04} in the Newtonian limit.

The mass-flux ratio $dM/d\psi$, which determines $F(\psi)$ via \eqref{F_gp_gr}, is determined itself by the history of the accretion process. In general, the accretion process is complicated, time-dependent, and nonaxisymmetric, as revealed by simulations \citep{romanova2015Accretion}. Identifying the flux surfaces that connect magnetically to the accretion disk, and calculating the instantaneous mass accretion rate on those flux surfaces, is an unsolved problem. In this paper, we follow previous authors \citep{pm04, priymak2011Quadrupole, suvorov2019Relaxation} and make the approximation
\begin{equation}
    \label{M_psi}
    M(\gp)=\frac{M_a}{2}\frac{1-e^{-\gp/\gp_a}}{1-e^{-\gp_*/\gp_a}},
\end{equation} 
with $\psi_\ast = \psi(R_*,\pi/2)$. In \eqref{M_psi}, $\psi_{\rm a}$ denotes the level surface which touches the inner edge of the accretion disk. Implicitly, we assume that $\psi_{\rm a}$ does not change during the accretion process. Equation \eqref{M_psi} then states that mass accretes mostly on the polar flux tube $0\leq \psi \leq \psi_{\rm a}$, i.e.\ along open magnetic field lines which connect magnetically to the accretion disk. Correspondingly, there is minimal accretion along equatorial magnetic field lines $\psi_{\rm a} \leq \psi \leq \psi_\ast$, which close inside the inner edge of the accretion disk. A nonzero amount of equatorial accretion is included to avoid a sudden density step at $\psi = \psi_{\rm a}$ and promote the numerical convergence of the Grad-Shafranov solver (see Section \ref{sec:numerics}), irrespective of whether or not it occurs astrophysically.

\subsection{Boundary conditions}
\label{sec:bc}

We consider the distortion of an initially dipolar magnetic field, which remains dipolar at and below the surface even after accretion, in the customary line-tying approximation \citep{pm04}. The general form of the flux function for a dipole is
\begin{equation}
    \label{gen_dipole}
    \psi_d(r,\theta)=\psi_* R_* g(r)\sin^2\theta.
\end{equation}
In the Newtonian theory, one has
\begin{equation}
    \label{rad_dip_nw}
    g(r)=\frac{1}{r}.  
\end{equation}
In general relativity, one has \citep{petterson1974Magnetic}
\begin{equation}
    \label{rad_dip_gr}
    g(r) = -\frac{3}{8}\frac{r^2}{M_*^3}\paren{\ln(1-\frac{2M_*}{r})+\frac{2M_*}{r}+\frac{2M_*^2}{r^2}}.
\end{equation}
Note that in both treatments, $\psi_{\rm d}$ is proportional to $\sin^2\theta$. However, in Newtonian theory one has $0<\psi_d(R_*,\theta)<\psi_*$, and in general relativity one has $0<\psi_d(R_*,\theta)<\psi_*R_*g(R_*)$, where $g(r)$ is given by \eqref{rad_dip_gr}. Additionally, the relativistic dipole \eqref{rad_dip_gr} satisfies the homogeneous relativistic Grad-Shafranov equation, that is, $\Delta^*\psi_d=0$.  

At $r=R_*$ the magnetic field is anchored in the heavy crust of the star, and we apply line-tying boundary conditions, that is, $\psi(R_*,\theta)=\psi_d(R_*,\theta)$. We also apply a Dirichlet condition along the magnetic axis, where we set $\psi(r,0)=0$, i.e., the polar field line remains straight after accretion. North-south symmetry at the equator also requires the magnetic field to be perpendicular to the $\theta=\pi/2$ plane, which translates to the Neumann boundary condition $\partial\psi/\partial\gt=0$ for all $r$ at $\theta=\pi/2$.

We restrict the numerical solver to the region $R_\ast \leq r \leq R_{\rm max}$ and choose $R_{\rm max}$ to lie outside most of the screening currents in the mountain, i.e.\ $R_{\rm max} \gtrsim 10^5$ mountain scale heights. At $r=R_{\rm max}$, we demand that the magnetic moment 
\begin{equation}
    \label{dip_mom_bc}
    m_{\rm d}(r)=\frac{3}{2}\frac{1}{g(r)}\int_{-1}^{1}\gp(r,\theta)\dd(\cos\theta),
\end{equation}
calculated by integrating $\psi$ over angle at fixed $r$, does not change as a function of $r$. This is justified physically, because the surface $r=R_{\rm max}$ is chosen deliberately to lie well outside the screening currents at $r\approx R_\ast$. Upon differentiating \eqref{dip_mom_bc} with respect to $r$, we obtain
\begin{equation}
    \label{strong_bc}
    0=\frac{3}{2}\frac{1}{g(r)}\int_{-1}^{1}\colch{-\frac{g'(r)}{g(r)}\gp+\pdv{\gp}{r}}\dd(\cos\theta).
\end{equation}
At large distances we expect that $\psi$ takes the form of a dipole ($\psi\propto\sin^2\theta$) as for any localized current distribution. This means, that the condition \eqref{strong_bc} can only be satisfied if one has
\begin{equation}
    \label{bc_outer}
    0=-\frac{g'(r)}{g(r)}\gp+\pdv{\gp}{r}
\end{equation}
for arbitrary large $r$. Equation \eqref{bc_outer}, evaluated at $r=R_{\rm max}$, provides a Robin-type boundary condition on $\psi$ to supplement the mixed Dirichlet and Neumann conditions specified above. 

\cite{pm04} chose the magnetic field to be radial at the outer boundary for numerical convenience. A radial field is inconsistent with the fact that any static current distribution tends to a dipole at large $r$, but the error thereby introduced is small, because the field is weak at $r=R_{\rm max}$, and the screening currents reside at $r \ll R_{\rm max}$, as confirmed numerically \citep{pm04}.

\section{Numerical scheme}
\label{sec:numerics}
We solve equations \eqref{iso_grad_shaf_eqn} and \eqref{F_gp_gr} simultaneously using a modified version of the iterative numerical scheme developed by \cite{pm04}. We review the algorithm and performance, with specific emphasis on the relativistic modifications, in this section. Other effects, such as superconductivity, have been studied by other authors and are omitted here to focus on relativistic corrections \citep{sur2021Impact}.

The algorithm presented here has been written in Python, to take advantage of inbuilt libraries for contouring and polynomial fitting, alongside \verb"numpy" vectorisation \citep{harris2020array} to optimise matrix operations such as those in the relaxation scheme described below. 
%brief summary
%grid
\subsection{Grid and dimensionless variables}
We solve equation \eqref{iso_grad_shaf_eqn} for $\psi(r,\theta)$ on a grid of $(N_r, N_{\theta})$ cells in $(r,\theta)$ coordinates, in the region $R_* \leq r \leq R_{\mathrm{max}}$ and $0 \leq \theta \leq \pi/2$. In keeping with \cite{pm04}, we convert to dimensionless coordinates $\Tilde{x} = (r - R_*)/x_0$ and $\Tilde{\mu} = \cos \theta$, and we introduce the dimensionless variables $\Tilde{\psi} = \psi/\psi_0$, $\Tilde{M} = M/M_a$, $\Tilde{F} = x_0^3 F/c_s^2 M_a$, $\Tilde{s} = s/x_0$, $x_0 = \frac{c_{\rm s}^2R_*^2}{GM_*}\alpha$, $\alpha = 1-\frac{2GM_*}{c^2R_*}$, $\beta = 1+\frac{c_{\rm s}^2}{c^2}$, and $a = R_*/x_0$.
% \begin{align}
%     \label{x0}
%     x_0 &= \frac{c_{\rm s}^2R_*^2}{GM_*}\alpha
%     \\
%     \label{alpha}
%     \alpha &= 1-\frac{2GM_*}{c^2R_*}
% \end{align}
In these variables, using the Grad-Shafranov operator defined in equation \eqref{grad_shaf_op_coords}, equations \eqref{iso_grad_shaf_eqn} and \eqref{F_gp_gr} become
\begin{align}
    \Tilde{\Delta}^*\Tilde{\gp}& =- x_0^4(\tilde{x}+a)^2(1-\tilde{\gm}^2)\dv{\tilde{F}}{\tilde{\gp}}e^{-\gb\tilde{x}}
    \label{non_dim_GS_eqn}
 \end{align}
 and
 \begin{align}
    \Tilde{F}(\Tilde{\psi}) &= \left(\frac{d\Tilde{M}}{d\Tilde{\psi}}\right)^{\beta}\left(2\pi\int_{C}(\Tilde{x}+a)(1-\Tilde{\mu}^2)^{1/2}|\Tilde{\nabla}\Tilde{\psi}|^{-1}e^{-\Tilde{x}}\dd{\Tilde{s}}\right)^{-\beta}
    \label{non_dim_F}
\end{align}
where the dimensionless Grad-Shafranov operator $\Tilde{\Delta}^{*}$ takes the form:
\begin{equation}
    \Tilde{\Delta}^{*} =\alpha e^{2(\beta-1)\Tilde{x}}\paren{\pdv[2]{}{\Tilde{x}}+2(\beta-1)\pdv{}{\tilde{x}}}+\frac{1-\tilde{\tilde{\gm}}^2}{(\tilde{x}+a)^2}\pdv[2]{}{\tilde{\gm}}
    \label{non_dim_gs_operator}
\end{equation}
To capture the large gradients in $\rho$ and $\psi$ near the surface of the neutron star, we space radial grid points logarithmically with maximum grid resolution nearest the surface according to $\Tilde{x}_1=\log\left(\Tilde{x} + e^{-L_x}\right) + L_x$, where the user-selected control parameter $L_x$ is chosen small (or zero) to ensure there are several grid points per scale height.
%boundary conditions
%mass-flux distribution - done
%solving F, curve-fitting, dF/dpsi analytic fit - done
%Poisson eqn - successsive over-relaxation - done
%convergence criteria - done
%verification/testing
\subsection{Integral mass-flux constraint}
To solve equations \eqref{non_dim_GS_eqn} and \eqref{non_dim_F}, we start with $dM/d\psi$ and an initial guess $\psi^{(0)}(r,\theta) = \psi_*R_*g(r)\sin^2\theta$. Contours of $\psi^{(0)}(r,\theta)$ are computed using the Python library \verb"contourpy". We choose $N_c = N_r - 1$ contours such that the contours and grid points are spaced comparably and therefore roughly optimally, as validated by \cite{pm04}.
With the contours in hand, $F[\psi^{(0)}]$ is calculated from equation \eqref{non_dim_F}. The derivative $F^{\prime}[\psi^{(0)}]$ is then computed from the polynomial fit:
\begin{equation}
    F(\psi) = \sum_{i=0}^{N_p}a_i \psi^i
    \label{numerical_polyfit}
\end{equation}
with $N_p = 8$. Polynomial fitting is performed using \verb"numpy.polynomial"'s \verb"polyfit" package, and analytic differentiation is performed by the \verb"polyder" package. As highlighted by \cite{pm04}, a polynomial fit avoids numerical instabilities introduced by simple finite-differencing. 
\subsection{Grad-Shafranov solver}
The right-hand side of equation \eqref{non_dim_GS_eqn} can be calculated after mapping $F^{\prime}[\psi^{(0)}]$ to the grid using bilinear interpolation. Equation \eqref{non_dim_GS_eqn}, which is an elliptic partial differential equation with a known source term on the right-hand side is then solved using successive over-relaxation to obtain the intermediate Gauss-Seidel iterate $\psi_{\mathrm{new}}^{(0)}$. The next iterate is obtained by under-relaxation, viz. $\psi^{(k+1)} = \Theta^{(k)}\psi^{(k)} + \left[1-\Theta^{(k)}\right]\psi_{\mathrm{new}}^{(k)}$, with under-relaxation factor $0 \leq\Theta^{(k)}\leq 1$. In order to speed up the relaxation scheme, we allow the relaxation parameter $\Theta^{(k)}$ to vary as the scheme progresses. We increase $\Theta^{(k+1)}$ towards unity when the residual $\delta^{(k)}$ defined by equation \eqref{eq:residual} at iteration $k$ is larger than the residual at iteration $k-1$, and decrease the relaxation parameter otherwise, according to
\begin{equation}
\Theta^{(k+1)} = 
    \begin{cases}
    \frac{\Theta^{(k)}+1}{2} &\delta^{(k)} > \sigma \delta^{(k-1)}\\
    2\Theta^{(k)} - 1 &\delta^{(k)} < \sigma \delta^{(k-1)}
    \end{cases}
\end{equation}
where $\sigma$ is a tolerance which prevents the relaxation parameter updating unless the residual changes between iterations by a factor $\sigma$. We choose $\sigma = 5$.

\subsection{Convergence}
\begin{figure*}
    \centering
    \begin{subfigure}{0.45\textwidth}
        \includegraphics[width=\textwidth]{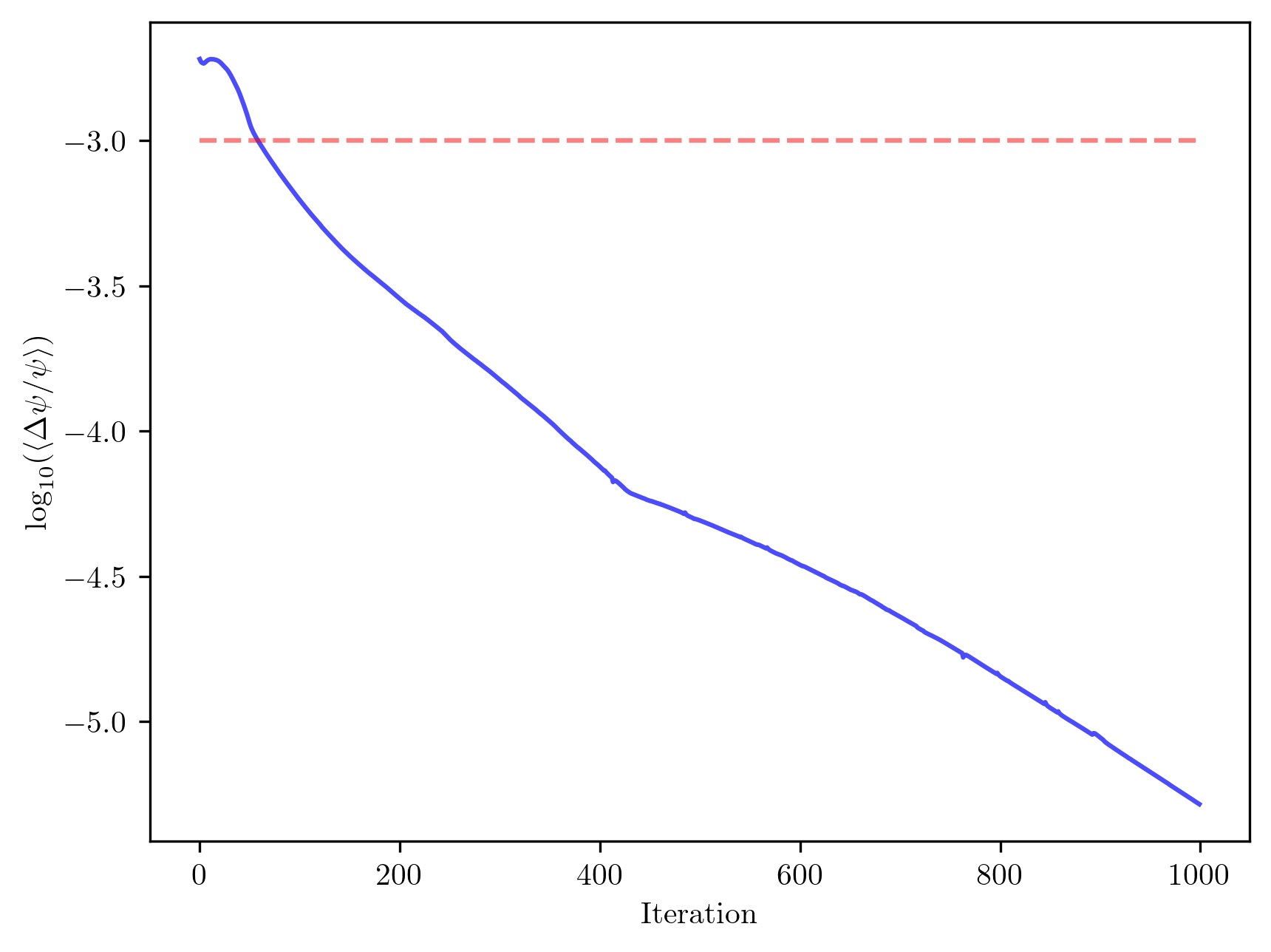}
        \caption{}
        \label{fig:residual}
    \end{subfigure}
    \begin{subfigure}{0.45\textwidth}
        \includegraphics[width=\textwidth]{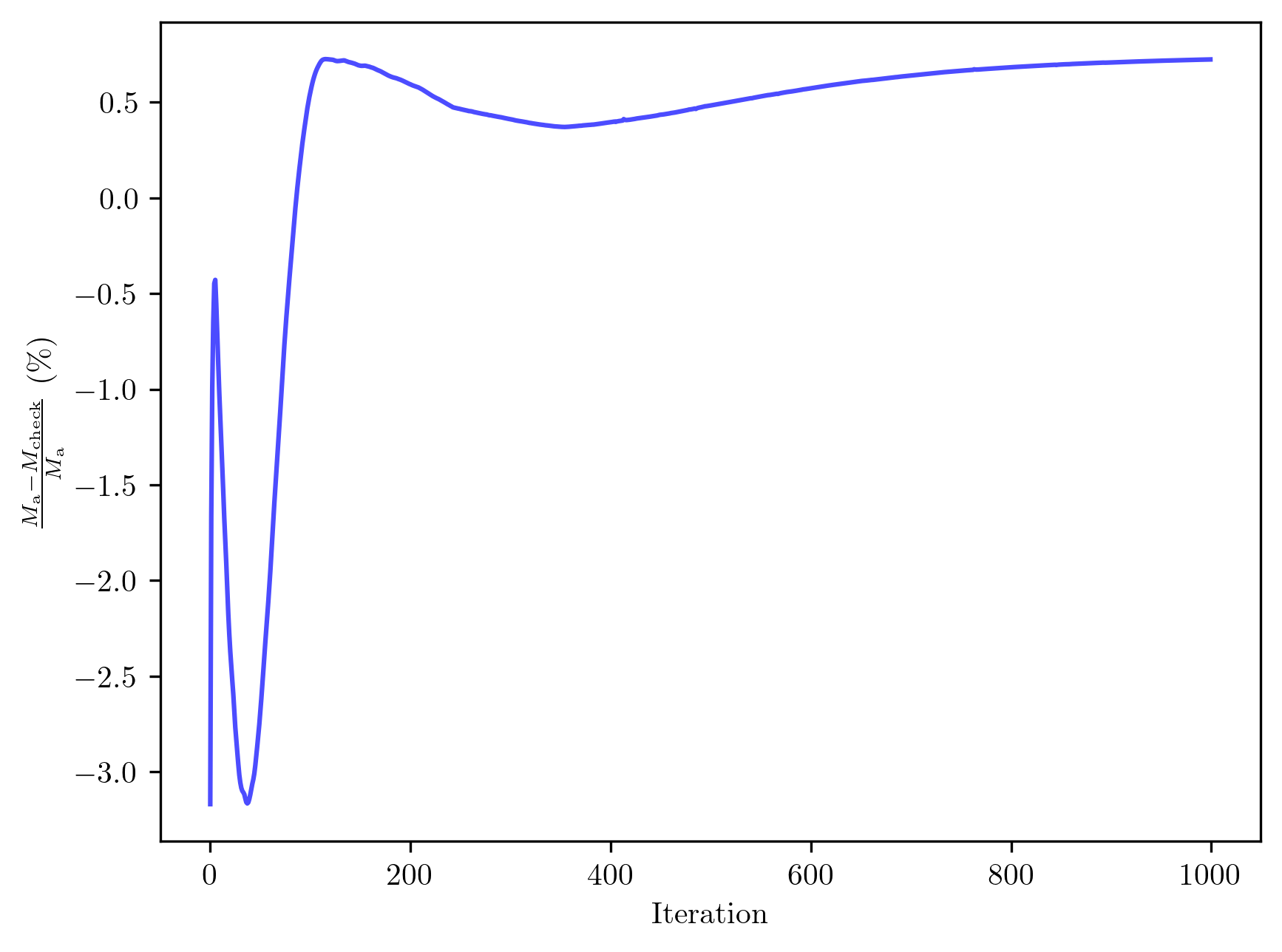}
        \caption{}
        \label{fig:mass_error}
    \end{subfigure}
    \caption{Convergence and validation test. (a) Residual $\left<\Delta\psi/\psi \right>$ versus iteration index $k$. (b) Fractional mass deficit $(M_{\rm a}-M_{\mathrm{check}})/M_{\rm a}$ versus $k$ as a percentage. Here, $\psi_*=5\times10^{23}\,\mathrm{G\,cm^2}$, $R_* = 10^6\,\mathrm{cm}$, $M_* = 1.4M_\odot$, $\Theta = 0.99$, $M_{\rm a} = 10^{-5}M_{\odot}$,  $b=10$, $N_r = N_\theta=256$, $N_{\rm c} = 255$.}
    \label{fig:combined_residuals}
\end{figure*}

Convergence of the solution is validated in two ways. Firstly, the mean residual over the grid is calculated viz.
\begin{equation}
    \left<\frac{\Delta\psi}{\psi}\right>^{(k)} = \frac{1}{N_r N_{\theta}}\sum_{i,j}\frac{|\psi^{(k)}(x_i, \mu_j) - \psi^{(k-1)}(x_i, \mu_j)|}{|\psi^{(k)}(x_i, \mu_j)|}
    \label{eq:residual}
\end{equation}
and iteration continues until the convergence criterion $\left<\Delta\psi/\psi \right>^{(k)}<\xi$ is satisfied. We choose $\xi = 10^{-3}$ typically. Figure \ref{fig:residual} displays the mean residual as a function of the iteration number. It can be seen that for $M_a = 10^{-5}M_\odot$, the solution converges within the first 100 iterations.
We additionally track the enclosed mass in the simulation domain, integrating equation \eqref{rho} over the volume $r \geq R_\ast$, which should be equal to the total accreted mass $M_a$. In practice we require that the enclosed mass be within approximately 5\% of $M_a$ to ensure there is minimal mass leakage. To check that this is the case, the density $\rho(\tilde{x},\mu)$ is integrated over the domain at each iteration to calculate the enclosed mass $M_{\mathrm{check}}$, defined by
\begin{align}
    \label{eq:checkmass_01}
    M_{\mathrm{check}} &= \int_V \rho(r,\theta,\phi)e^{-\Phi}r^2\sin\theta \dd{r} \dd{\theta} \dd{\phi} \\
    &= \frac{2\pi x_0^3}{\sqrt{\alpha}}\int_0^{\tilde{x}_{max}}\int_0^{1}(\tilde{x}+a)^2 \rho(\tilde{x},\mu)e^{-(\beta-1)\tilde{x}}\dd{\mu} \dd{\tilde{x}}.
\end{align}
Converting to a discrete sum to allow summation over the grid, and multiplying by a factor of two to account for both hemispheres, we obtain
\begin{equation}
    M_{\mathrm{check}} = \frac{4\pi x_0^3}{\sqrt{\alpha}}\sum_{i=0}^{N_{{x}}}\sum_{j=0}^{N_{\mu}} (\tilde{x}+a)^2 \rho[x_i,\mu_j] e^{-(\beta-1)\tilde{x}}\Delta\mu\Delta{x}.
    \label{eq:checkmass_02}
\end{equation}
Figure \ref{fig:mass_error} displays the percentage difference between the total accreted mass and the enclosed mass, illustrating that there is no significant mass leakage from our simulation domain. One might ask why $|M_{\rm a} - M_{\rm check}|$ does not tend to zero, as $k$ increases. This occurs because of the imperfect spatial resolution of the numerical simulation, i.e. the mass discrepancy is due to the nonzero grid size, not the finiteness of $k$.

\section{Mountain properties}
\label{sec:results}
In this section, we present numerical solutions of the relativistic Grad-Shafranov equation \eqref{iso_grad_shaf_eqn} subject to the flux freezing condition \eqref{mass-flux}. We have verified that every relativistic result smoothly goes to the correspondingly Newtonian result when we artificially increase the speed of light in our code ($c\mapsto2c$, $4c$, $8c$ and $10c$). The magnetic field profiles and the mass density of the mountain are discussed in Section \ref{sec:hydromag_structure}. The magnetic dipole moment is calculated as a function of $M_{\rm a}$ in Section \ref{sec:mag_burial}. In every plot in this section that displays ``altitude'', the latter quantity is defined to be the proper radial distance measured by static observers. Furthermore, in every simulation, we adopt the following fiducial values of the physical parameters: $M_*=1.4 M_{\odot}$, $R_*=10^6\,\mathrm{cm}$, $\psi_*=5\times10^{23}\,\mathrm{G\,cm^2}$ and $c_s=10^8\,\mathrm{cm\, s^{-1}}$, copying \cite{pm04} to facilitate comparison.

\subsection{Relativistic hydromagnetic structure}
\label{sec:hydromag_structure}

We start by investigating how the hydromagnetic structure of the mountain, with general relativistic corrections considered, depends on $M_{\rm a}$ and $b$ in the ranges $10^{-6} \leq M_{\rm a} \leq 3\times10^{-5}$ and $b=3$ or $10$, in keeping with the Newtonian analysis \citep{pm04}. The parameter $b$ can be interpreted geometrically in terms of the colatitude $\theta_{\rm a} = \arcsin(b^{-1/2})$, where the flux surface $\psi_{\rm a}$ in the undistorted dipole ($M_{\rm a} = 0$) meets the stellar surface, i.e. the half-opening-angle of the magnetic polar cap.

\begin{figure*}
    \centering
    \begin{subfigure}{0.4\textwidth}
        \includegraphics[width=\textwidth]{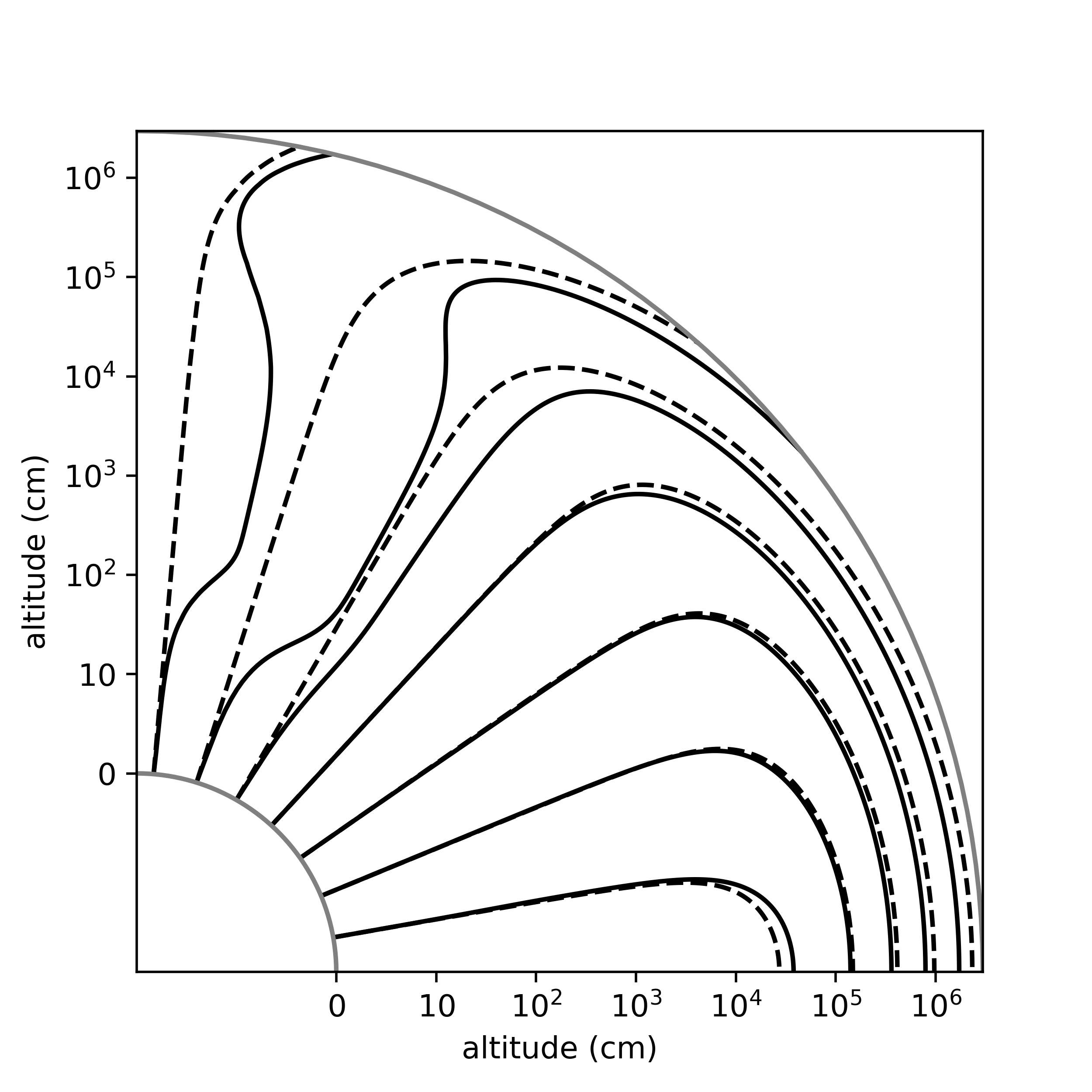}
        \caption{}
        \label{fig:gr_and_dipole_psi}
    \end{subfigure}
    %\hfill
    \begin{subfigure}{0.4\textwidth}
        \includegraphics[width=\textwidth]{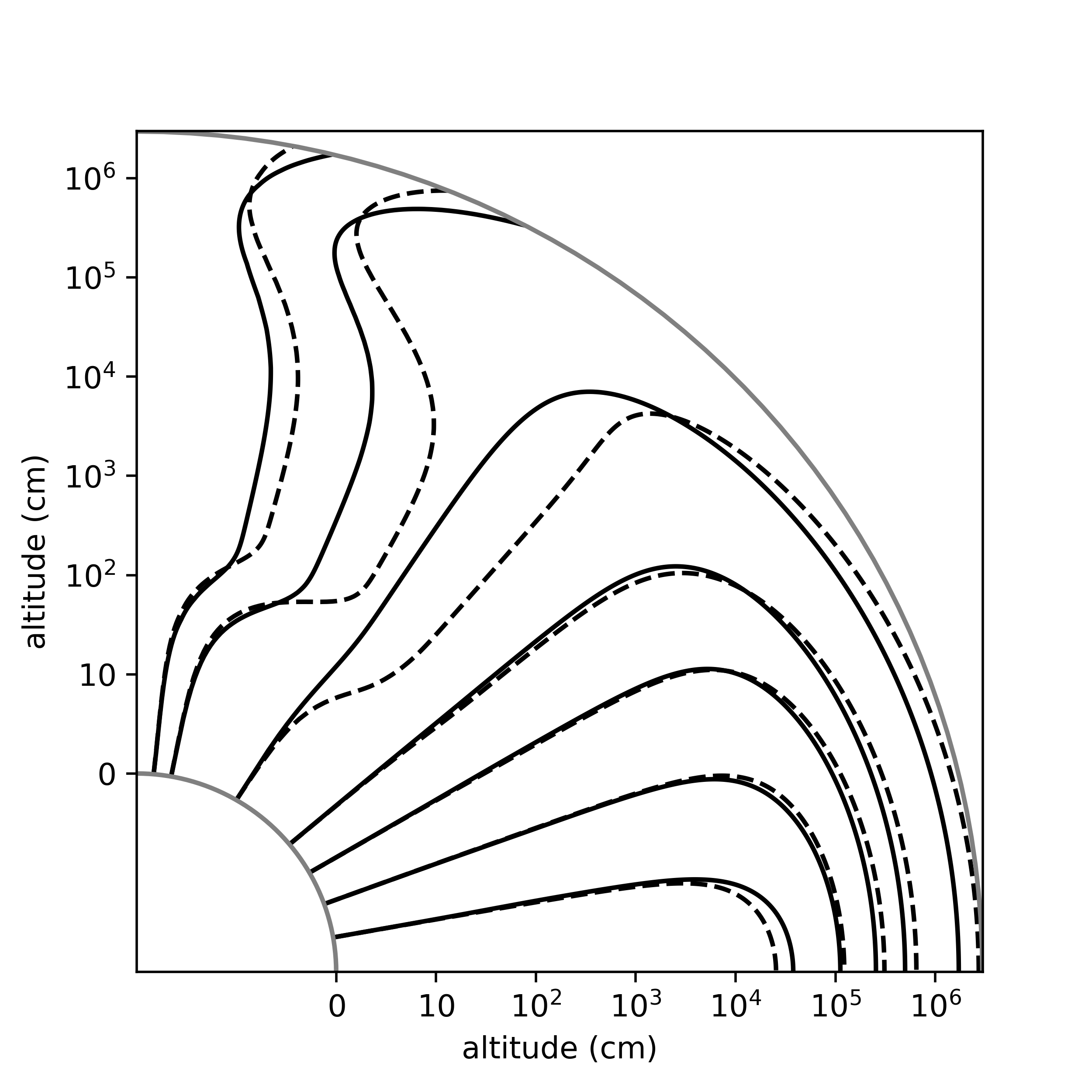}
        \caption{}
        \label{fig:gr_and_nw_psi}
    \end{subfigure}
    \caption{Flux function $\psi$ for $M_a=10^{-5}M_\odot$ and $b=10$. (a) Final relativistic configuration (solid lines) compared to the initial dipole (dashed lines). (b) Final configuration in relativistic (solid lines) and Newtonian (dashed lines) treatments.}
    \label{fig:psi}
\end{figure*}

Figure \ref{fig:gr_and_dipole_psi} displays contours of the flux function $\psi$ (in cross-section at fixed longitude) as solid curves. The contours are for the representative values $b = 3$ and $M_{\rm a} = 10^{-5}M_\odot$ such that the mountain distorts the magnetic field significantly. In the same panel, we overlay the initial undistorted dipole given by \eqref{gen_dipole} and \eqref{rad_dip_gr}, depicted as dashed curves. The result is qualitatively the same as in \cite{pm04}, i.e., the polar mountain deforms the field lines into a ``magnetic tutu'' concentrated around the equator, and the polar mountain is confined by the $\theta$ component of the magnetic tension in the equatorial tutu. 

Figure \ref{fig:gr_and_nw_psi} compares directly the final equilibrium $\psi$ for the relativistic and Newtonian scenarios, drawn with solid and dashed contours respectively. To perform this comparison, we renormalise the contours of the Newtonian case by multiplying $\psi$ by $R_*g(R_*)$, where $g(r)$ is given by \eqref{rad_dip_gr}. It is evident that general relativity introduces significant changes to the flux function. The magnetic deformation is not as pronounced, i.e. the curvature of the magnetic field lines is lower, in the relativistic treatment. This aligns with the expectation discussed at the end of the Section \ref{sec:theory}.

\begin{figure*}
    \centering
    \begin{subfigure}{0.45\textwidth}
        \includegraphics[width=\textwidth]{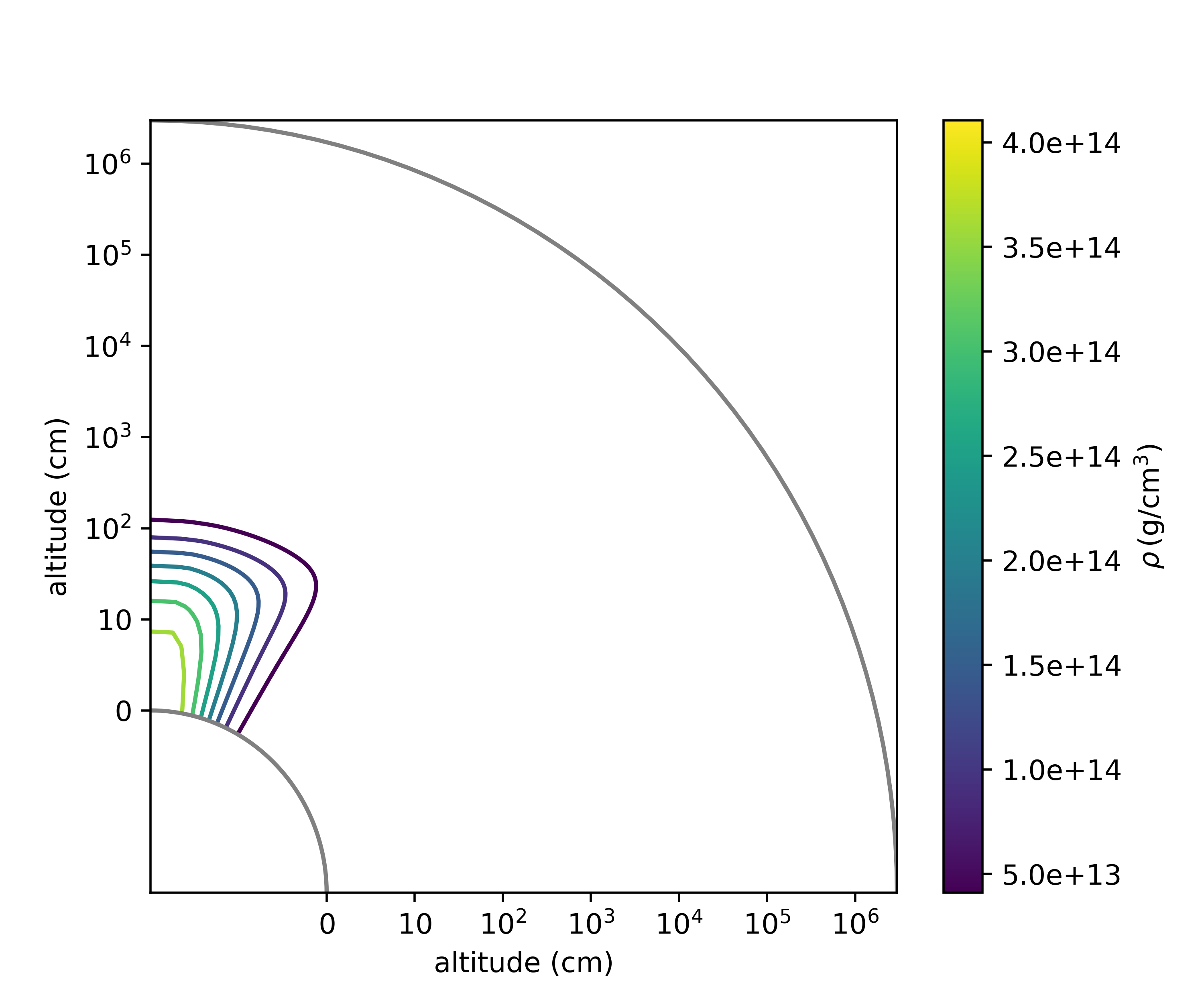}
        \caption{}
        \label{fig:density_gr}
    \end{subfigure}
    %\hfill
    \begin{subfigure}{0.45\textwidth}
        \includegraphics[width=\textwidth]{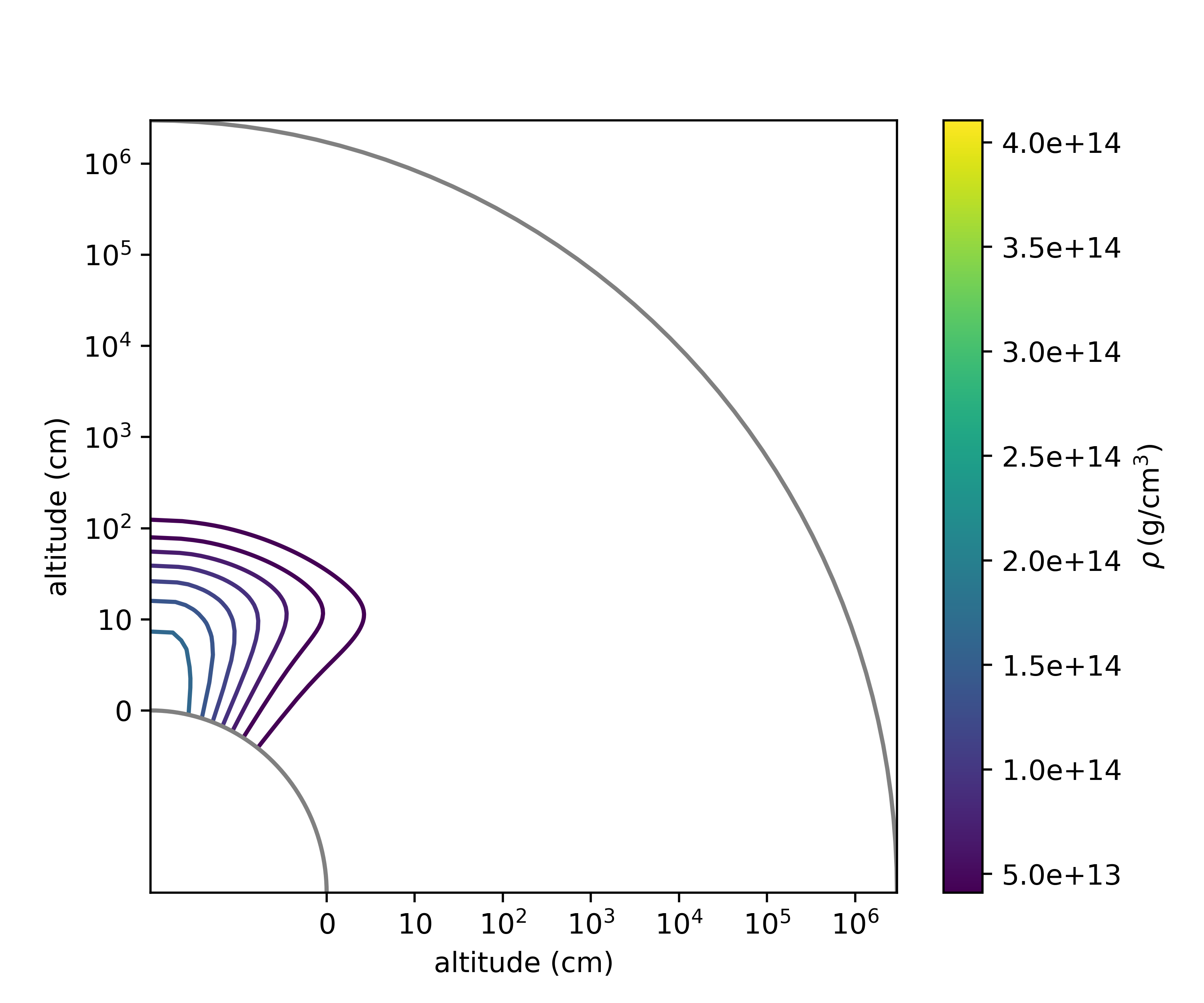}
        \caption{}
        \label{fig:density_nw}
    \end{subfigure}
    %hfil
    \begin{subfigure}{0.45\textwidth}
        \includegraphics[width=\textwidth]{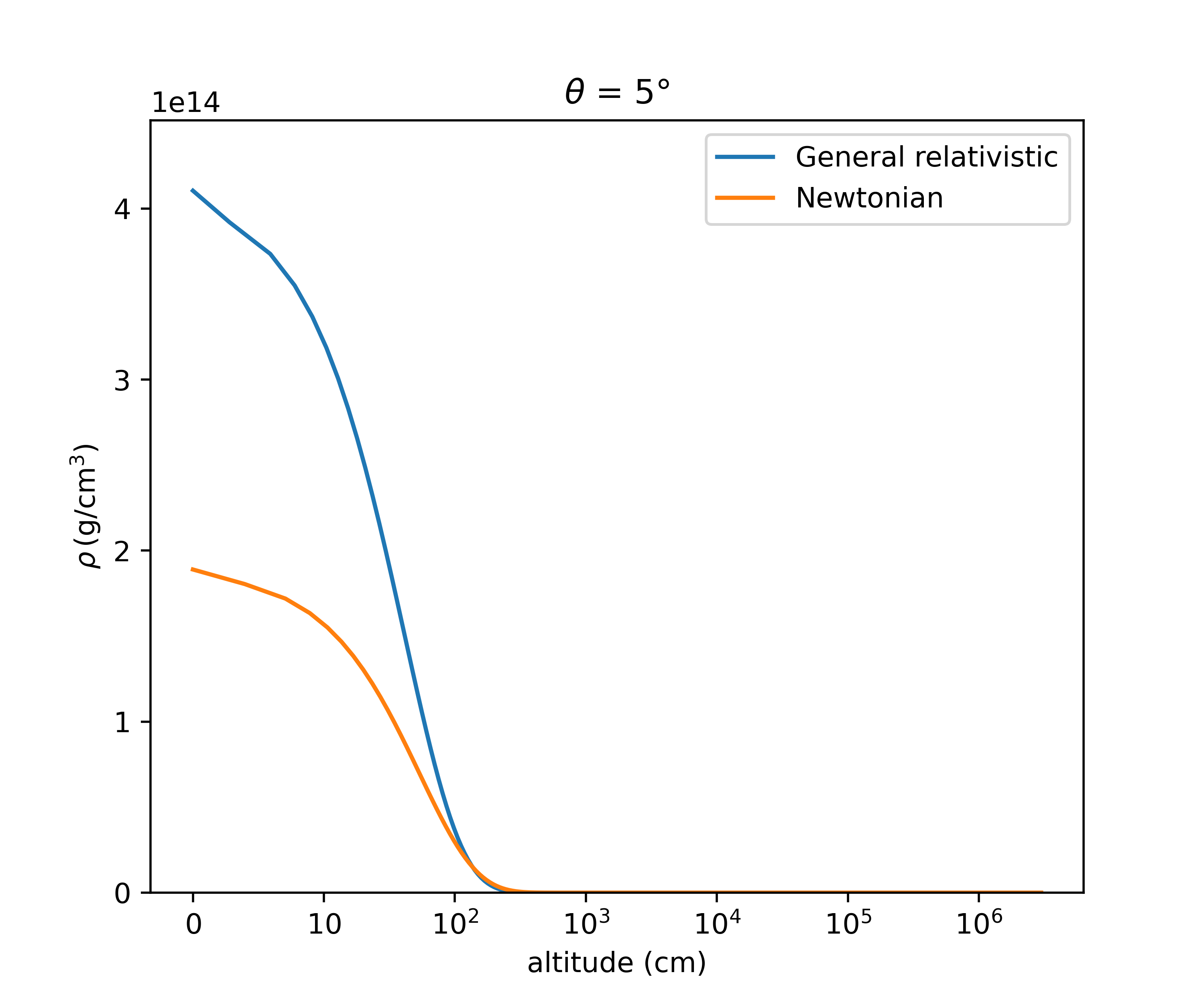}
        \caption{}
        \label{fig:density_per_r}
    \end{subfigure}
    %hfil
    \begin{subfigure}{0.45\textwidth}
        \includegraphics[width=\textwidth]{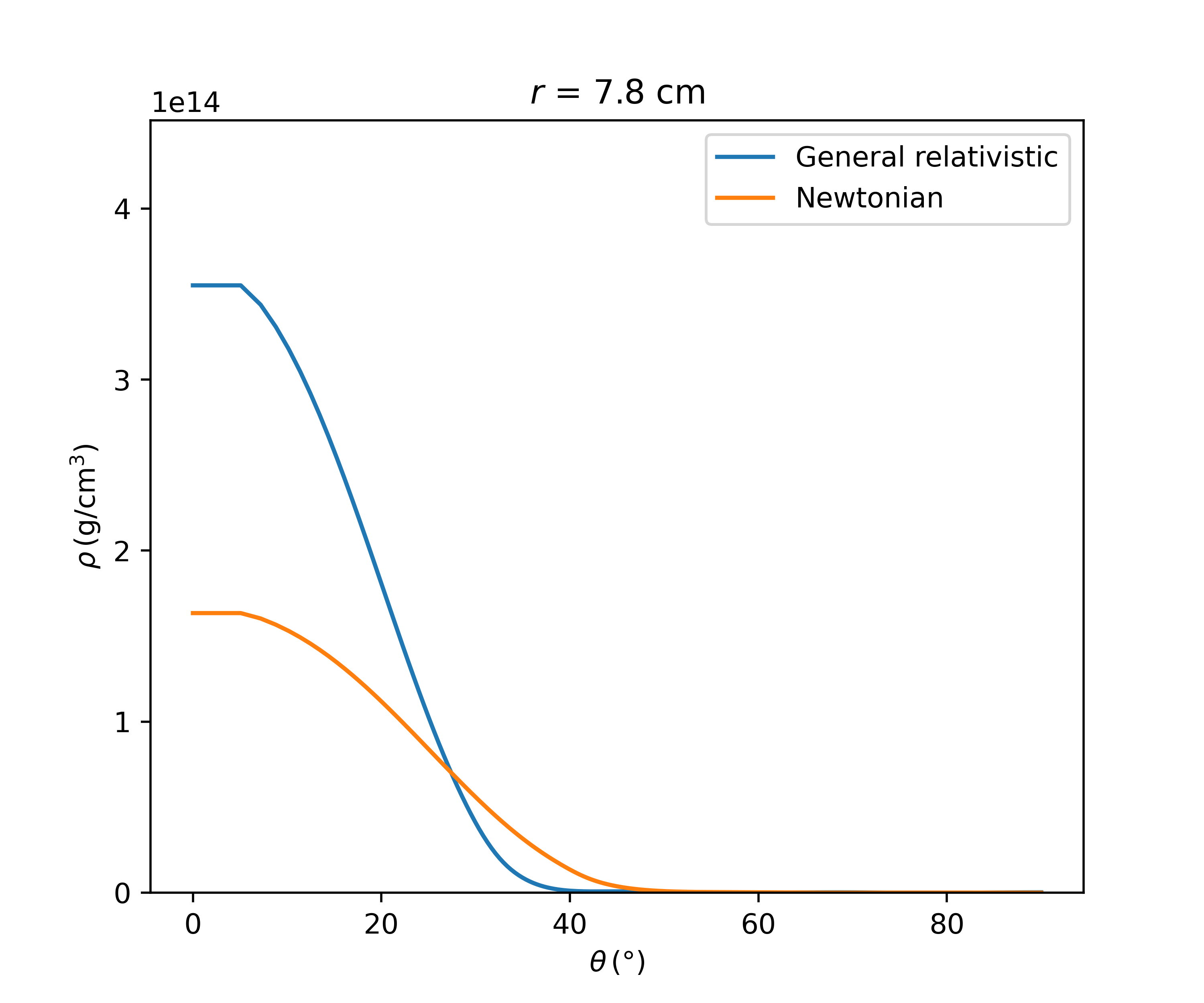}
        \caption{}
        \label{fig:density_per_theta}
    \end{subfigure}
    \caption{Density configuration of the magnetically confined mountain. Panels (a) and (b) display the contour levels of the density $\rho$ in the general relativistic and Newtonian scenarios, respectively. Panel (c) shows the density profile in cross-section at $\theta= 5\degree$ in both treatments. Panel (d) shows the density profile in cross-section at a constant altitude ($7.8\,\rm{cm}$).}
    \label{fig:density_gr_vs_nw}
\end{figure*}

Figure \ref{fig:density_gr_vs_nw} complements Figure \ref{fig:psi} by displaying the density structure of a mountain with $M_{\rm a} = 10^{-5}$. Panels \ref{fig:density_gr} and \ref{fig:density_nw} show the density contours with and without general relativistic corrections, respectively. The maximum density when considering general relativity is $4.1\times10^{14}\,\rm{g\,cm^{-3}}$, compared to $1.9\times10^{14}\,\rm{g\,cm^{-3}}$ in the Newtonian theory. These maxima of density are reached on the surface of the star at the magnetic poles. As discussed in \cite{pm04}, they are unrealistically high due to the rigid surface assumption. A complete treatment of this problem should include the sinking of the mountain into the crust, as studied by \cite{wette2010Sinking}.

In panel \ref{fig:density_per_r}, we fix the colatitude at $\theta = 5\degree$ and plot $\rho$ against altitude. Although the curves differ by a factor greater than two at the surface, they have the same asymptotic behaviour for altitude $\gtrsim10^2\,\rm{cm}$. This behaviour is observed across all colatitudes. Panel \ref{fig:density_per_theta} shows the colatitudinal dependence of the density at the fixed altitude of $7.8\,\rm{cm}$. We notice that the mountain is confined nearer to the pole in the relativistic case. In panel \ref{fig:density_per_theta}, the two curves cross at $\theta\approx25\degree$. Similar behavior is observed for other altitudes inside the mountain.

\subsection{Magnetic burial: reduction of the magnetic dipole moment}
\label{sec:mag_burial}

The magnetic dipole moment $m_{\rm d}$ in general relativity is a global quantity calculated by integrating the magnetic flux over a volume containing the electric currents of interest. In order to compare with the Newtonian results reported previously in the literature, including by \cite{pm04} and \cite{priymak2011Quadrupole}, we generalize the Newtonian formula for $m_{\rm d}$ to encapsulate both the Newtonian and the relativistic regimes according to
\begin{equation}
    \label{dip_mom}
    m_{\rm d}(r)=\frac{3}{2}\frac{1}{g(r)}\int_{-1}^{1}\gp(r,\gm)\dd\gm
\end{equation}
Equation \eqref{dip_mom} determines the dipole moment at a fixed $r$, that is, it incorporates the effect of the diamagnetic screening currents in the volume $R_\ast \leq r' \leq r$ but not the screening currents in the volume $r' \geq r$. Equation \eqref{dip_mom} gives the Newtonian formula when we choose $g(r)$ to be given by equation \eqref{rad_dip_nw}.

Caution should be exercised when comparing Newtonian and general relativistic values of $m_{\rm d}$. Equation \eqref{B_from_F} shows that the magnetic field depends on the observer who measures it, and equation \eqref{rad_dip_gr} shows that the form of the dipole is not the same in both theories. This leads to subtleties in setting the value of $\psi_*$ in our simulations. A way to approach this would be to fix a fiducial value of the polar magnetic field measured by a local observer in both theories and then compute the appropriate value of $\psi_*$. A second approach is to fix a fiducial value for $\psi_*$ and to use that value in both theories. The latter approach avoids the conversion of $B_*$ to $\psi_*$ and makes the simulations agree for the value of the dipole moment calculated with equation \eqref{dip_mom} at the surface. Given these helpful properties, we choose the latter approach. 

\begin{figure*}
    \centering
    \begin{subfigure}{0.45\textwidth}
    \includegraphics[width=\textwidth]{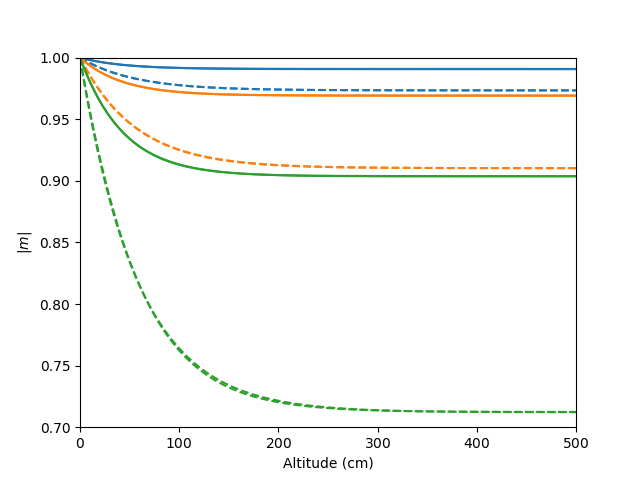}
        \caption{}
        \label{fig:dipole_per_r}
    \end{subfigure}
    \begin{subfigure}{0.45\textwidth}
    \includegraphics[width=\textwidth]{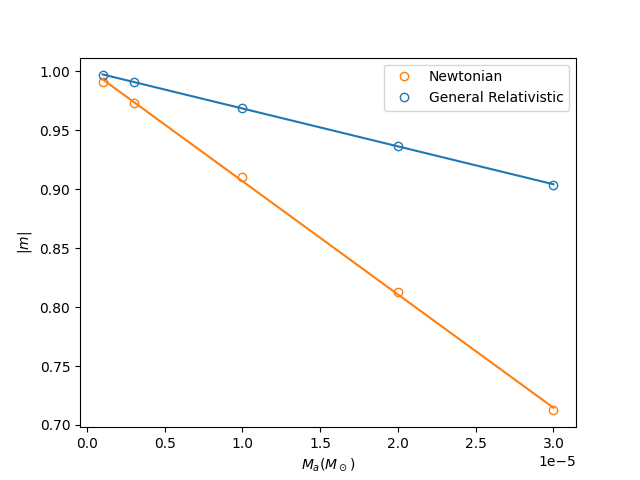}
        \caption{}
        \label{fig:dipole_per_Ma}
    \end{subfigure}
    \caption{Normalized magnetic dipole moment. (a) Magnetic burial for different masses in the general relativistic (solid lines) and Newtonian (dashed lines) scenarios. The different accreted masses are colour coded as follows: $10^{-6}M_\odot$ in blue, $10^{-5}M_\odot$ in orange, and $3\times10^{-5}M_\odot$ in green. (b) Normalised dipole moment as a function of the accreted mass $M_a$ for Newtonian (orange circles) and general relativistic  (blue circles) treatments, overlaid with the best fit linear regressions. For all these simulations, we take $b=3$.}
    \label{fig:dipole_mom}
\end{figure*}

In Figure \ref{fig:dipole_mom}, we show how the normalized magnetic dipole moment depends on the altitude and on the accreted mass. The normalisation is done with respect to the pre-accretion dipole moment. Panel \ref{fig:dipole_per_r} shows the decay of the normalised magnetic dipole moment as a function of altitude in both the Newtonian (as in \cite{pm04}) and general relativistic scenarios. We choose $b=3$ and three different values of accreted mass: $10^{-6} M_{\odot}$, $10^{-5} M_{\odot}$ and $3\times10^{-5} M_{\odot}$. Qualitatively, screening occurs near the surface in both scenarios. Nevertheless, the screening currents are compressed to lower altitudes in the relativistic scenario. Performing an exponential fit of the form
\begin{equation}
    \label{exp_fit}
    y=(1-y_\infty)e^{-x/\lambda}+y_\infty,
\end{equation}
to the data in Figure \ref{fig:dipole_per_r}, we find $\lambda = 42\,\mathrm{cm}$ (general relativity), versus $\lambda = 55\,\mathrm{cm}$ (Newtonian), approximately independent of $M_{\rm a} \leq 3\times 10^{-5} M_\odot$. The fitted $\lambda$ values are comparable to the scale heights $x_0 = 32 \, {\rm cm}$ (general relativity) and $x_0 = 54 \, {\rm cm}$ (Newtonian).

Figure \ref{fig:dipole_per_Ma} shows the decrease of the normalised dipole moment as a function of accreted mass for both the Newtonian and general relativistic theories. In the same figure, we plot the linear regressions:
\begin{align}
	\label{lin_reg_gr}
	m_{\rm d,GR} = -3.21\times10^{3}M_{\rm a}/M_\odot + 1.00
	\\
	\label{lin_reg_nw}
	m_{\rm d,N} = -9.61\times10^{3}M_{\rm a}/M_\odot + 1.00
\end{align}
for the general relativistic and Newtonian scenarios respectively. The linear regressions \eqref{lin_reg_gr} and \eqref{lin_reg_nw} show that $m_{\rm d}$ is around three times higher in the relativistic scenario for the same $M_{\rm a}$.

\section{Conclusion}
\label{sec:conclusion}
We have verified that general relativistic corrections are important in the theory of magnetically confined mountains on accreting neutron stars. The main effect of relativity is to smooth out the deformation of the magnetic field. This smoothing results in a three-fold decrease in the screening effects of the magnetic field burial. We also conclude that the length scale of the mountain is $40\%$ smaller when compared to the Newtonian result.

In our analysis, we make several approximations that can be relaxed in future work. These approximations include: perfect conductivity \citep{vigelius2009Resistive}, isothermal equation of state \citep{priymak2011Quadrupole, suvorov2019Relaxation}, neglect of type II superconductivity \citep{passamonti2014Quasiperiodic, sur2021Impact} and Schwarzschild spacetime, which does not include contributions to the metric from the rotation of the star. Another interesting future investigation would be to evolve the equations dynamically in a fully general relativistic magnetohydrodynamic simulation. This would test the stability of the equilibrium solutions that we calculate \citep{vigelius2008Threedimensional, mukherjee2012Phasedependent, mukherjee2013MHD} and also appropriately describe the interaction of the mountain with the spacetime geometry. This latter effect is likely to be small for the mountain sizes dealt with in this paper, but could become more relevant as one increases the accreted mass to values comparable to that of the star.

Finally, the results presented here have consequences for the generation of gravitational waves \citep{melatos2005Gravitational}. The amplitude of the persistent, quasi-monochromatic signal emitted by a magnetic mountain will be investigated in a forthcoming paper.

\section*{Acknowledgements}
We thank Arthur Suvorov, Maxim Priymak, and Donald Payne for providing access to the Grad-Shafranov solver upon which we have based the solver used in this paper, and for their assistance in the setup of the original code.

This work received financial support from the Catalyst Fund provided by the New Zealand Ministry of Business, Innovation and Employment and administered by the Royal Society Te Apārangi; the Division of Science of the University of Otago, New Zealand; the University of Otago Postgraduate Publishing Bursary (Doctoral); and the Australian Research Council Centre of Excellence for Gravitational Wave Discovery (OzGrav), through project number CE170100004.

\section*{Data availability}
The data underlying this article will be shared on reasonable request to the corresponding author.

\bibliographystyle{mnras}
\bibliography{GRMHD_Paper}

%%%%%%%%%%%%%%%%%%%%%%%%%%%%%%%%%%%%%%%%%%%%%%%%%%

%%%%%%%%%%%%%%%%% APPENDICES %%%%%%%%%%%%%%%%%%%%%

\appendix

%\section{Differential Forms and Exterior Calculus}
%\label{app:forms}
%\input{Sections/appendixA}

\section{Symmetries of Magnetically confined mountains}
\label{app:sym}
Continuous symmetries of a system are expressed mathematically by Lie derivatives. The connection between exterior and Lie derivatives is given by \textit{Cartan's Identity},
\begin{equation}
    \label{cartan_idn}
    \liedv{X}F=X\cdot\dd{F}+\dd{(X\cdot F)},
\end{equation}
with the dot symbolising the contraction of the vector $X$ with the first index of the form $F$.

Since our system is axisymmetric, the Lie derivative of the Faraday tensor with respect to the Killing vector $e_\gf^a$ is zero. Therefore, by the relation \eqref{cartan_idn}, we have
\begin{equation}
    \dd{(e_{\gf}\cdot F)}=0,
\end{equation}
where we use Maxwell's equation \eqref{max_eqn1}. Since the exterior derivative of $e_{\gf}\cdot F$ vanishes, according to Poincaré's Lemma, there is a function $\gp$ satisfying
\begin{equation}
    \label{psi_origin}
    e_{\gf}^aF_{ab}=\codv{b}\gp.
\end{equation}

By virtue of equation \eqref{psi_origin}, $\codv{a}\gp$ is normal to $e_{\gf}^a$. Furthermore, from the ideal MHD condition $F_{ab}u^b=0$, we have that $\codv{a}\gp$ is normal to $u^a$ and, hence, to $e_{t}^{a}$. Given all the above conditions, we can write the most general form of $F_{ab}$ as
\begin{equation}   F=\dd{\gf}\wedge\dd{\gp}+C\star(\dd{\gf}\wedge\dd{\gp}).
\end{equation}

If the magnetic field has zero toroidal component, i.e. $(\star F)_{ab}e^b_{\gf}=0$, we have $C=0$. Therefore, we can write the electromagnetic tensor as
\begin{equation}
    \label{F_sym}
    F=\dd{\gf}\wedge\dd{\gp}.
\end{equation}

Equation \eqref{F_sym} shows that the Faraday tensor $F$ is completely determined from the function $\psi$. That means that every magnetic quantity can be calculated once $\psi$ is known.

%\section{Comparing Newtonian and Relativistic Formalisms}
%\label{app:nw_vs_gr}
%\input{Sections/appendixC}

%%%%%%%%%%%%%%%%%%%%%%%%%%%%%%%%%%%%%%%%%%%%%%%%%%

% Don't change these lines
\bsp	% typesetting comment
\label{lastpage}
\end{document}